\documentclass[aps,prb,twocolumn,groupedaddress,showpacs]{revtex4}  

\usepackage{graphics}
\usepackage{epsfig}  

\begin{document}  
  
\title{  
Electronic properties of structural twin and antiphase   
boundaries in materials with strong electron-lattice couplings  
}  
  
\author{K. H. Ahn \cite{Ahn}, T. Lookman, A. Saxena, and A. R. Bishop}   
\affiliation{Theoretical Division, Los Alamos National Laboratory,   
Los Alamos, New Mexico 87545}  
  
  
\begin{abstract}  
  
Using a symmetry-based atomic scale theory of lattice distortions,   
we demonstrate that elastic textures, such as twin and antiphase boundaries,  
can generate intricate electronic inhomogeneities in materials with   
strong electron-lattice  
coupling, as observed in perovskite manganites and other functional materials.  
  
\end{abstract}  
 
 
\maketitle  
  
\newpage  
Recent advances in imaging techniques  
have revealed the presence of rich elastic textures   
in functional materials such as  
colossal magnetoresistive (CMR) manganites,\cite{Uehara99,Fukumoto99}  
ferroelectrics,~\cite{Abakumov00} ferroelastics,\cite{Cao01} 
and shape memory alloys.\cite{Ren02}  
In particular, recent experiments on  
certain perovskite manganites compounds\cite{Fukumoto99}  
have shown the correlation between   
electronic transport properties and the presence of meandering  
antiphase boundaries (APBs)   
within insulating charge ordered domains,  
interpreted as the existence of metallic regions forming around APBs.  
It is also reported that strains near grain boundaries  
in thin film can modify electronic properties in manganites.\cite{Soh00}   
The interplay between elastic texture and   
electronic heterogeneity   
is thus central to understanding      
multiphase coexistence and the resultant electronic properties 
 in CMR and other functional materials.   
  
In this Report,   
we illustrate the importance of elastic inhomogeneities  
in modifying electronic properties in materials with strong electron-lattice coupling.   
In particular, we  study the electronic properties of APBs and twin boundaries (TBs) on a  
two-dimensional (2D) lattice.  
We first show how a recently-developed symmetry-based atomic-scale   
theory of lattice distortions  
can be used to find  
atomic configurations of twin and antiphase   
boundaries.  
Within our framework, we illustrate the differences 
and similarities between TBs and APBs from the point of view of  
localization of long [short] wavelength modes inside 
APBs [TBs],  evolution upon energy relaixtion and  
roughness~\cite{TB.v.APB} [smoothness] of APBs [TBs].  
We subsequently perform a tight binding calculation  
with a Su-Schrieffer-Heeger (SSH) type model of electron-lattice coupling  
to predict the distribution of electronic density of states,   
which can be related to the results  
of scanning tunneling microscope (STM) measurements.   
Our work thus forms the basis for predicting  
electronic properties from predesigned materials microstructures.  
  
\begin{figure}  
\leavevmode  
\epsfxsize6.5cm\epsfbox{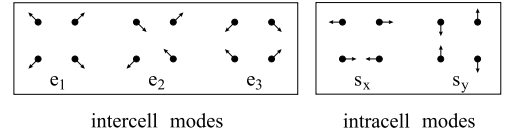}  
\caption{\label{fig:modes}   
Distortion modes for   
a square lattice in 2D with a monatomic basis.\cite{Ahn03}  
For example, $s_x(\vec{i})=[d^x(\vec{i})-d^x(\vec{i}+(10))+  
d^x(\vec{i}+(11))-d^x(\vec{i}+(01))]/2$,  
where $d^x(\vec{i})$ and $d^y(\vec{i})$ represent the displacement  
of the atom at site $\vec{i}$ along $x$ and $y$ directions respectively.  
}  
\end{figure}

Twin boundaries separate domains related by the rotation of crystalline axes,  
whereas APBs represent boundaries at which the sequence   
of alternating distortions,   
such as alternating rotational directions of oxygen octahedra  
in perovskite oxides, change their phase  
(i.e., broken translational symmetry).   
Although our method can be applied to 2D or 3D lattices  
with mono- or multiatomic bases,   
we illustrate our ideas with a square lattice in   
2D space with   
a monatomic basis,   
for which the appropriate atomic scale distortion variables are   
the modes shown in Fig.~\ref{fig:modes} (Ref.~\onlinecite{Ahn03}).  
These modes have important advantages over displacement variables --  
they reflect the symmetries of the lattice and   
can serve as order parameters (OP) in structural phase transitions.  
Therefore, energy expressions with desired ground states  
can be written in a simpler way  
in these variables than in usual displacement variables.  
Moreover, since the lattice distortions are decomposed into   
the modes at $\vec{k}=(0,0)$ (long-wavelength or intercell modes)   
and $\vec{k}=(\pi,\pi)$ (short-wavelength or intracell modes),  
the approach using these modes shows the differences between  
long and short wavelength lattice distortions in a natural way.  
  
\begin{figure}  
\leavevmode  
\epsfxsize3.0cm\epsfbox{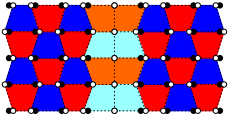}  
\caption{\label{fig:apb.disp}  
(Color) An example of the atomic displacement pattern on either side of an APB.   
Solid and open circles represent the atomic positions for undistorted square  
lattice  
and distorted lattice, respectively.  
Red and blue represent the positive and negative signs of the $s_x$ mode  
defined for each plaquette.  
}  
\end{figure}

We consider APBs,  
such as the one shown in  
Fig.~\ref{fig:apb.disp}  
where open circles represent the distorted atomic positions,  
for $s_x$ or $s_y$ modes.  
The simplest energy expression yielding a ground state with  
either pure $s_x$ or $s_y$ mode lattice distortion  
is   
\begin{eqnarray}  
E_{\text{sxsy}}&=&\sum_{\vec{i}} - \frac{D}{2} [ s_x(\vec{i})^2 + s_y(\vec{i})^2 ]   
+\frac{G_1}{4} [ s_x(\vec{i})^4 + s_y(\vec{i})^4 ]  
\nonumber \\  
&+&   
\frac{G_2}{2}  s_x(\vec{i})^2 s_y(\vec{i})^2 +   
\sum_{\vec{i},n=1,2,3} \frac{C_n}{2} e_n(\vec{i})^2, \label{eq:Esxsy}  
\end{eqnarray}  
where the coefficients $D$, $G_1$, $G_2$ (with $G_1<G_2$) can be obtained  
from interatomic forces and $C_n$ are the associated elastic moduli.  
  
Since the ground state has  $\vec{k}=(\pi,\pi)$ component lattice distortions,  
we define variables with tilde by multiplying $(-1)^{i_x+i_y}$ with    
all intercell/intracell/displacement variables  
[e.g., $\tilde{s}_x (\vec{i})$=$s_x (\vec{i}) (-1)^{i_x+i_y}$,  
$\tilde{d}_x (\vec{i})$=$d_x (\vec{i}) (-1)^{i_x+i_y}$].  
In wavevector space, this corresponds to interchanging   
the $(0,0)$ and $(\pi,\pi)$ points.  
Therefore,   $\tilde{s}_x$ and $\tilde{s}_y$ are the modes near   
$\vec{q}=(0,0)$,   
and $\tilde{e}_1$, $\tilde{e}_2$, and $\tilde{e}_3$ are the modes  
near $\vec{q}=(\pi,\pi)$,   
where $\vec{q}$ represents the wavevector for the variables with tilde.  
The constraint equations~\cite{Ahn03}  
in the continuum limit  
are  
$  
\tilde{e}_{1,3}(\vec{r}) = (\nabla_y \tilde{s}_x \pm \nabla_x \tilde{s}_y)/(2\sqrt{2})  
$ and   
$  
\tilde{e}_{2}(\vec{r}) = (\nabla_x \tilde{s}_x + \nabla_y \tilde{s}_y)/(2\sqrt{2})   
$.  
These relations show that   
spatial variations of {\it intracell} modes   
($\tilde{s}_{x}$, $\tilde{s}_{y}$)  
always generate {\it intercell} modes ($\tilde{e}_{1}$, $\tilde{e}_{2}$, $\tilde{e}_{3}$),  
which becomes microscopic origin   
of the phenomenological gradient energy term in continuum elasticity theory.  
Therefore,  
intercell modes are present inside an APB at which   
an intracell mode with tilde changes its sign.  
Similar arguments apply to a TB~\cite{Ahn03} --   
the spatial variations of intercell modes   
always generate intracell modes inside TB through their constraints --  
and  lead to a similar solitary profile   
for TB and APB.~\cite{Barsch84,Cao01}  
However, there exists a fundamental difference between intercell and   
intracell modes:   
In the $k \rightarrow 0$ limit, $e_1$, $e_2$, and $e_3$ are  
given as a first derivative of $d_x$ and $d_y$,   
whereas in the $q \rightarrow 0$ limit,   
$\tilde{s}_x$ and  $\tilde{s}_y$ are $2\tilde{d}_x$ and $2\tilde{d}_y$.  
Therefore, $e_1$, $e_2$, and $e_3$ are related by  constraint equations,  
whereas $\tilde{s}_x$ and  $\tilde{s}_y$ are not constrained   
by each other.  
  
We consider the APB between the two ground states  
with $ \tilde{s}_x (\vec{i}) = \pm \sqrt{D/G_1} $  
(Fig.~\ref{fig:apb.disp}).  
On expanding Eq.~(\ref{eq:Esxsy}) around the ground state and retaining leading order   
terms, we write the energy   
$E_{\text{APB}}=E_{\text{APB,OP}}+E_{\text{APB,NOP}}$, where  
\begin{eqnarray}  
E_{\text{APB,NOP}}&=&\sum_{\vec{i},n=1,2,3} \frac{C_n}{2} \tilde{e}_n(\vec{i})^2  
+\sum_{\vec{i}} \frac{D'}{2}\tilde{s}_y(\vec{i})^2  \\  
E_{\text{APB,OP}}&=&\sum_{\vec{i}}  
-\frac{D}{2}\tilde{s}_x(\vec{i})^2+\frac{G_1}{4}\tilde{s}_x(\vec{i})^4,   
\end{eqnarray}   
and $D'=D(G_2/G_1-1)$.  
The energy $E_{\text{APB}}$ has the identical form to   
that for   
a TB with $e_3$ as OP~\cite{Ahn03} --  
a double-well potential for OP and harmonic potentials for non-OP.  
For both TB and APB problems,   
the non-OP energy terms mediate an anisotropic   
interaction between OPs through the constraints.  
The minimization of $E_{\text{APB,NOP}}$ using  
Lagrange multipliers leads to  
$E_{\text{APB,NOP}}^{\text{min}}=\sum_{\vec{q}}   
\frac{1}{2} \tilde{s}_x(-\vec{q}) \tilde{U}(\vec{q}) \tilde{s}_x(\vec{q})$,  
where $\tilde{U}(\vec{q})$ has a $q^2$ leading order term with an anisotropic coefficient   
$\tilde{U}_2(\theta_q) = [ (C_1+C_2+C_3)+(C_2-C_1-C_3) \cos 2\theta_q ]/16$.  
When transformed into real space,  
the $q^2$ leading order term gives rise to a short range $R^{-4}$ interaction   
between OP $\tilde{s}_x$,  
where $R$ is the distance between two sites,  
unlike the long-range $R^{-2}$ interaction between OP $e_3$   
for the TB case.\cite{Shenoy99,LongRange}  
Such different range of interaction is consistent with  
well-known understanding\cite{TB.v.APB} of smoothness of TB and roughness of APB,  
which our simulation reproduces below.  
  
The physical origin of the long-range interaction between {\it intercell} modes  
and  short-range interaction between {\it intracell} modes  
lies in the difference in the symmetry operations  
relating the two domains separated by a TB or APB.  
For a TB, the sign change in OP corresponds to  
a change in orientation, which has no intrinsic length scale  
and thus gives rise to  a long range interaction.  
For an APB, the sign change in OP signifies translation of the configuration  
by one atomic spacing,  
implying the presence of an   
intrinsic atomic length scale that is responsible for the fast decay of   
the interaction between OPs.  
These considerations apply    
to any 2D/3D lattice with mono- or multiatomic bases.  
We can also expect it to be valid for   
the rough and fluctuating stripes suggested in high-$T_C$ cupurates,  
which are examples of magnetic APBs.  
Arguments similar to those in Ref.~\onlinecite{Ma.SM},  
related to the range of the interactions,  
explain the characteristic smoothness and roughness   
of TBs and APBs.  
  
\begin{figure}  
\leavevmode  
\epsfxsize8.5cm\epsfbox{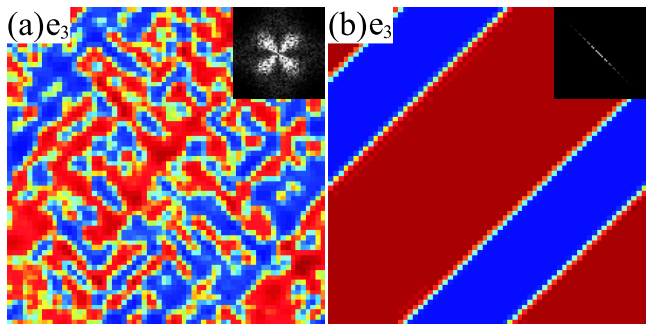}  
\caption{\label{fig:tb.sim}   
(Color) $e_3(\vec{i})$ (main panels) and   
$|e_3(\vec{k})|$ (insets)   
obtained from  simulation for TB:  
(a) early and (b) late stages of the lattice relaxation.  
Parameter values (see Ref.~\onlinecite{Ahn03} for the definitions)  
are $A_1=A_2=B=4$, $A_3=5$, and $F_3=50$.  
The center of the inset corresponds to $\vec{k}=0$ and   
the four corners $\vec{k}=(\pm\pi,\pm\pi)$.     
(Dark red: 0.32, dark blue: -0.32,  
white color in the inset:  
larger than 0.2 times the maximum of $|e_3(\vec{k})|$).  
}  
\end{figure}

\begin{figure}  
\leavevmode  
\epsfxsize8.5cm\epsfbox{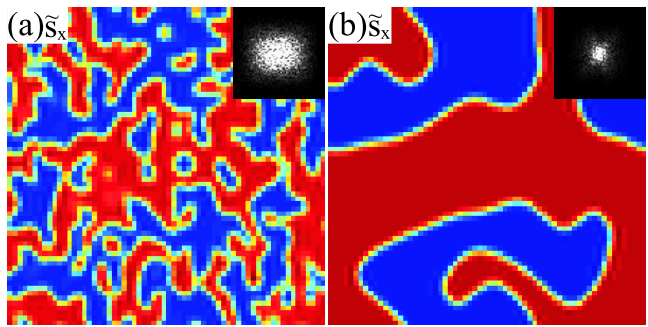}  
\caption{\label{fig:apb.sim}   
(Color) $\tilde{s}_x(\vec{i})$ (main panels) and   
$|\tilde{s}_x(\vec{q})|$ (insets) distributions  
obtained from  simulations for APB:  
(a) early and (b) late stage of the lattice relaxation.  
Parameter values are $C_1=C_2=C_3=D'=4$,  
$D=5$, and $G_1=50$.    
}  
\end{figure}

The full expression of the kernel $\tilde{U}(\vec{q})$ in $E_{\text{APB,NOP}}^{\text{min}}$  
is given by                             
\begin{eqnarray}  
\tilde{U}(\vec{q})&=&\frac{C_1 + C_3}{2} T_y(\vec{q})^2 +  
\frac{C_2}{2} T_x(\vec{q})^2 \\  
&-&\frac  
{(C_1+C_2-C_3)^2 T_x(\vec{q})^2 T_y(\vec{q})^2 }  
{ 4  D' +   
2 (C_1 + C_3) T_x(\vec{q})^2 + 2 C_2 T_y(\vec{q})^2 },  
\nonumber  
\end{eqnarray}  
where $T_x(\vec{q})=\tan (q_x/2)$ and $T_y(\vec{q})=\tan (q_y/2)$.  
We choose similar parameter values for both APB and TB cases    
and use the Euler method~\cite{Shenoy99}  
to relax the lattice  
starting from random initial conditions  
to see difference in evolution between TB and APB.  
The results, $e_3$ for TB and $\tilde{s}_x$ for APB,   
on a $64 \times 64$ lattice with  periodic boundary conditions  
are shown   
in Figs.~\ref{fig:tb.sim} and \ref{fig:apb.sim}, respectively,  
where the main panels show real space distributions and insets   
the $k$ or $q$-space distributions.  
Figures~\ref{fig:tb.sim}(a) and \ref{fig:apb.sim}(a) correspond to  
early stages of the relaxation,   
which show the characteristics of the OP distribution at high temperature($T$),  
whereas Figs.~\ref{fig:tb.sim}(b) and \ref{fig:apb.sim}(b)  
reflect the   
OP distributions for late stages at low $T$.  
Even at the early stage, the presence of the  
long range correlation between intercell OP along $45^o$ and $135^o$ can be identified in   
the main panel in Fig.~\ref{fig:tb.sim}(a), reminiscent of tweed structures   
in martensitic materials.~\cite{Shenoy99}  
The $k$-space distribution plotted in the inset of Fig.~\ref{fig:tb.sim}(a)  
shows strong preference of $e_3(\vec{k})$ along $45^o$ and $135^o$ orientations.  
Such long-range correlation and anisotropy are absent for   
the phase of the intracell mode distortion, as shown in Fig.~\ref{fig:apb.sim}(a).  
The late stage of the relaxation   
depicted in  Figs.~\ref{fig:tb.sim}(b) and \ref{fig:apb.sim}(b)  
show characteristic smooth TBs~\cite{Ren02}  and rough APBs~\cite{Fukumoto99}.   
The TB is metastable and cannot be removed by further relaxation,  
unless  large noise is applied.    
In contrast, the APB in Fig.~\ref{fig:apb.sim}(b),   
which is a ring due to the periodic boundary condition,  
shrinks and disappears upon further relaxation.  
Although the solitary-wave profile of the smooth APB   
along a certain direction is a metastable state,\cite{Cao01}   
the absence of a long range interaction between the phase   
of intracell mode distortions  
prevents the relaxation of random initial configuration  
from reaching such metastable states.  
This indicates that lattice defects or boundary conditions are necessary   
to reach the metastable solitary-wave APB configuration in materials,  
and  influence the geometry of APBs.

\begin{figure}  
\leavevmode  
\epsfxsize8.5cm\epsfbox{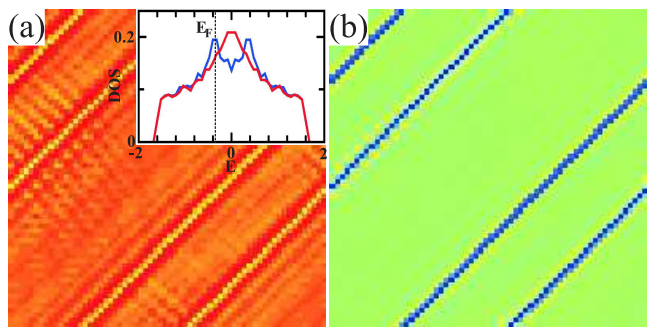}  
\caption{\label{fig:tb.e}   
(Color) Electronic properties calculated for the  
lattice distortion in Fig.~\ref{fig:tb.sim}(b).  
(a) Inset: local DOS within the domain (blue curve) and   
TB (red curve);  
main panel: spatial distribution of local DOS at $E_F$.  
(b) Corresponding charge density distribution  
(dark blue: 0.36, green: 0.39).   
}  
\end{figure}

\begin{figure}  
\leavevmode  
\epsfxsize8.5cm\epsfbox{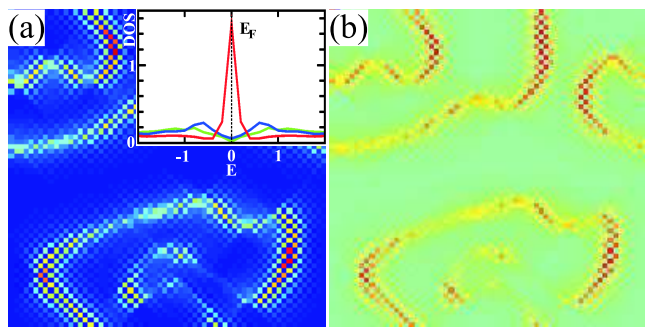}  
\caption{\label{fig:apb.e}   
(Color) Electronic properties calculated for the  
lattice distortion in Fig.~\ref{fig:apb.sim}(b).  
(a) Inset: local DOS within the domain (blue curve) and   
APB (red and green curves);  
main panel: spatial distribution of local DOS at $E_F=0$  
(dark blue: 0, dark red: 2).  
(b) Spatial charge density distribution for $E_F=0.1$  
(green: 0.52, dark red: 0.72).   
}  
\end{figure}

\begin{figure}  
\leavevmode  
\epsfxsize8.5cm\epsfbox{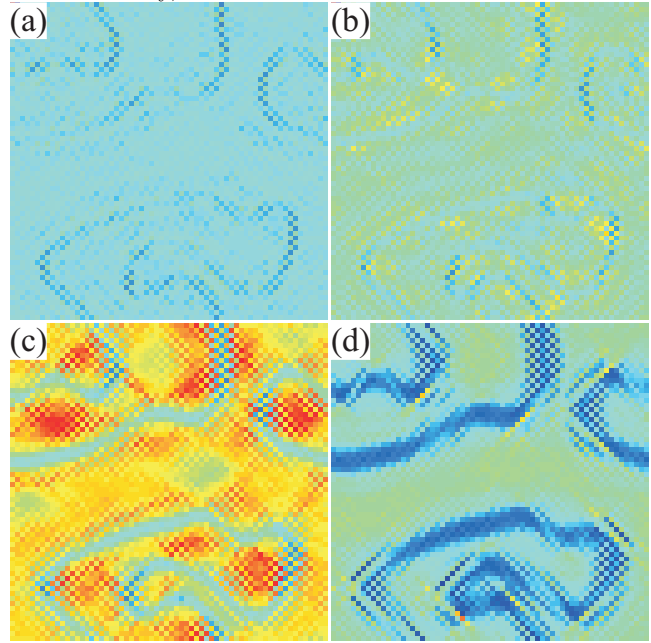}  
\caption{\label{fig:apb.ldos}   
(Color) Local DOS calculated for the  
lattice distortion in Fig.~\ref{fig:apb.sim}(b) at various $E_F$ values  
(a) -1.5, (b) -1.1, (c) -0.7, and (d) -0.3.  
(dark red: 0.35, dark blue: 0)   
}  
\end{figure}

To study  functional electronic aspects associated with these  
TB and APB microstructures, we consider  
the modulation of electronic properties  
based on the following   
SSH electron-lattice coupling Hamiltonian,  
\begin{equation}  
H_{\text{SSH}}=\sum_{\vec{i},a=x,y}   
-t_0 [1-\alpha (d^a_{\vec{i}+\hat{a}}-d^a_{\vec{i}})]   
(c^{\dagger}_{\vec{i}} c_{\vec{i}+\hat{a}} + c^{\dagger}_{\vec{i}+\hat{a}} c_{\vec{i}}),   
\end{equation}  
where $c^{\dagger}_{\vec{i}}$ is the creation operator for an electron  
at $\vec{i}$.  
We use $d^x(\vec{i})$ and $d^y(\vec{i})$ obtained from  
our atomistic model   
as inputs to the SSH Hamiltonian.  
For $t_0=1$, $\alpha=1$, and the TB and APB results shown in   
Figs.~\ref{fig:tb.sim}(b) and~\ref{fig:apb.sim}(b),  
we find all energy levels and eigenstates numerically,  
and calculate the local density of state (DOS)  
at each site and   
the distributions of local DOS    
and charge density  
for chosen Fermi energies ($E_F$),  
as shown in Figs.~\ref{fig:tb.e},~\ref{fig:apb.e} and \ref{fig:apb.ldos}.  
  
The blue curve in the inset of Fig.~\ref{fig:tb.e}(a)   
represents the local DOS within the domain of Fig.~\ref{fig:tb.sim}(b), 
close to the bulk DOS for the homogeneous phase. 
The energy difference between the two peaks   
is proportional to $\alpha |e_3|$.  
Within the TB, $|e_3|$ is small   
and therefore the local DOS spectral weight moves toward $E=0$,  
as shown with a red curve.   
This local shift of the DOS weight can be measured   
using direct local probes, such as  
STM.~\cite{Uehara99,Davis01}  
For $E_F$ shown in this inset, the local DOS is smaller inside the  
TB than inside the domain,  
and the main panel of Fig.~\ref{fig:tb.e}(a)  
shows  
the real space distribution of local DOS at this $E_F$.  
The oscillation of local DOS within the domain  
is related to Friedel oscillations.  
The feedback from electron to lattice neglected here  
may generate a similar Friedel oscillation of the lattice.  
The charge density, which is the area in local DOS below $E_F$,  
is a constant (0.5) if $E_F=0$.  
If $E_F<0$,  more electrons are depleted from the TB than the domain,  
as shown in Fig.~\ref{fig:tb.e}(b).  
The charge density also has a Friedel oscillation,  
though not clearly visible in Fig.~\ref{fig:tb.e}(b).  
  
The inset of Fig.~\ref{fig:apb.e}(a) shows  
the local DOS for the APB case.  
Within the domain, each site has a  neighbor along   
$x$ axis that is closer in one direction   
and farther in the other   
(Fig.~\ref{fig:apb.disp}).  
Such a distortion pattern results   
in a V-shaped local DOS  
with zero DOS at $E=0$,   
plotted as a blue curve in the inset of Fig.~\ref{fig:apb.e}(a)  
(the small finite value of local DOS at $E=0$   
is due to the finite energy bin size).  
Within the APB, the average of  
the two bond lengths along $x$-axis has a  
$(\pi,\pi)$ component. 
For example,  
along sites at the center of the $90^o$ APB shown in Fig.~\ref{fig:apb.disp},  
the neighboring sites in the $x$-direction  
are  alternately closer or farther apart. 
The red line in the inset of Fig.~\ref{fig:apb.e}(a)  
represents   
the local DOS for the sites with two farther neighbors  
within the center of APB. This   
has a peak near $E=0$.  
The green line represents the local DOS  
at sites with two closer neighbors,   
right next along $y$ direction to the sites for the red curve.   
The difference between the red and green curves emphasizes the difference  
that occurs in density of states  due to atomic-scale changes in  
microstructure as along an APB.  
Such atomistically sharp 
changes at structural interfaces  
have been seen in CMR manganites with STM   
(Renner {\it et al.}~\cite{Uehara99}).   
The local DOS at $E_F=0$ is zero within the domain  
and finite only around the APB with a $(\pi,\pi)$ component modulation,  
as depicted in the main panel of Fig.~\ref{fig:apb.e}(a).  
This means that the electronic states created by the APB  
dominate   
the low energy properties  
of the whole system, e.g.,   
conductivity or specific heat.  
The charge density for $E_F$ slightly higher than zero  
is plotted in Fig.~\ref{fig:apb.e}(b).  
Checkerboard-type modulations of electronic properties around APBs, 
reflecting the electronic states coupled to the  
$(\pi,\pi)$ type lattice distortions,  
demonstrate that STM studies of TB and APB can be used   
to reveal underlying electronic  
properties of materials (usually hidden in a homogeneous phase),  
similar to the way single impurity atoms have been used to reveal  
superconducting properties in high T$_C$ cupurates.  
Results for different values of $E_F$ are displayed in   
Fig.~\ref{fig:apb.ldos}.  
Local DOS modulations, particularly the checkerboard type modulations, 
tend to reach farther into the domain  
than the elastic texture itself.  
We also note that the checkerboard type modulations  
with a wave vector $(\pi,\pi)$ are present at any $E_F$, 
in contrast to Friedel oscillations for TBs 
which change wavevectors with $E_F$.   
 
Although the orbital states and type of electron-lattice coupling  
in CMR manganites are quite different from the simple model  
presented here, our results suggest that simple  
APBs from double-well type potentials, such as Eq.(3), 
may give rise to electronic heterogeneities, 
but not undistorted metallic regions  
with uniform charge densities near APBs. 
An energy landscape with a local minimum at undistorted state, 
such as the one considered in Ref.~\onlinecite{Ahn04}, 
would be more likely to nucleate metallic domains at APBs, 
and create percolating conducting paths in CMR manganites.  
 
In summary, we have shown that in functional materials with  
strong electron lattice coupling,  
the electronic properties   
are modified near elastic textures such as TBs and APBs, 
which can be directly measured by STM. 
The results also show that the heterogeneities of electron local DOS  
are not just confined within TBs and APBs,   
but can  propagate into domains  
in the form of Friedel oscillations for TBs  
and with the wave vector of short wave length lattice distortions  
for APBs.   
 
This work was supported by the US DOE.

\end{document}